# Data Reduction for Time-of-Flight Small Angle Neutron Scattering with Virtual Neutrons

DU Rong[1,2], TIAN HaoLai[1,2,*], ZUO TaiSen[1,2], TANG Ming[1,2], YAN LiLi[1,2],

ZHANG JunRong[1,2,*]

[1]*China Spallation Neutron Source, Institute of High Energy Physics,*

*Chinese Academy of Sciences, Dongguan 523803 (Beijing 100049), China*

[2]*Dongguan Institute of Neutron Science, Dongguan 523808, China*

[*]Corresponding authors
Tel.: +86 769 89254869
E-mail: jrzhang@ihep.ac.cn (ZHANG JunRong).
E-mail: tianhl@ihep.ac.cn (TIAN HaoLai).
Mailing address: No.1 Zhongziyuan Road, Dalang Town, Dongguan, Guangdong province, China. 523000
This work is supported by the National Natural Science Foundation of China (No. 11305191), and the Project of Key Laboratory of Neutron Physics, China Academy of Engineering Physics (No. 2014BB04).

**Abstract:**

Small-angle neutron scattering (SANS) is an experimental technique to detect material structures in the nanometer to micrometer range. The solution of the structural model constructed from SANS strongly depends on the accuracy of the reduced data. The time-of-flight (TOF) SANS data are dependent on the wavelength of the pulsed neutron source. Therefore, data reduction must be handled very carefully to transform measured neutron events into neutron scattering intensity. In this study, reduction algorithms for TOF SANS data are developed and optimized using simulated data from a virtual neutron experiment. Each possible effect on the measured data is studied systematically, and suitable corrections are performed to obtain high-quality data. This work will facilitate scientific research and the instrument design at China Spallation Neutron Source.

# Introduction

The small-angle neutron scattering (SANS)[1] technique is widely used to detect the structure of materials in the nanometer to micrometer range[2–3], and it has been applied in various fields including nanotechnology, biotechnology, environmental sciences, soft matter, and energy sciences[4–8]. SANS based on time-of-flight (TOF) technology[9] has been developed in spallation neutron sources[10-12]. In addition, time-of-flight grazing incidence small-angle neutron scattering (TOF-GISANS) can be performed with such an instrument, which makes it highly promising for various future applications[13-14]. The highly quality of the data is the key to obtain accurate results. Therefore, detailed and flexible data processing should be performed after the experiment.

Monte Carlo simulation is a powerful tool for the design and optimization of neutron scattering facilities and instruments[15]. A virtual experiment is a complete experiment that covers every element from the source to the detector with Monte Carlo simulations[16], and it is performed using several software programs[17-19]. McStas[20-21] is utilized to simulate a SANS virtual experiment in this study. The digitization and readout module of the data processing framework[22] developed for the China Spallation Neutron Source provides a standard interface between McStas simulating neutron instruments and the raw data. The readout module generates event-based binary files with the same format as measurement data files. The post-processing module that is also based on the data processing framework receives raw data from the simulation package, and it reconstructs the raw data into a NeXus[23] data file. The NeXus file serves as a container for all relevant data associated with a scientific instrument or beamline. It provides the standard format for data processing and further analysis.

The scattering data in a TOF-SANS instrument are stored as two-dimensional maps of neutron counts versus pixel identification of the detector with a certain neutron wavelength[24]. This histogram loses the time-stamping information when it is generated, and this can introduce an error in rebinning[25]. In order to improve data quality, some new methods were proposed. A smart strategy is taken at the REFSANS instrument at Heinz Maier-Leibnitz zentrum[26]. The data are acquired in list mode, all neutrons are saved with their time-of-flight, and wavelength conversion is performed later[27]. In addition, the TOF-SANS data at the spallation neutron source at Oak Ridge national laboratory are stored in accordance with the neutron hit event in the NeXus format, which is the universal data format used in neutron scattering experiments. Considering that a large number of neutron events exist in the NeXus data, which need to be transformed into neutron scattering intensity, data reduction is one of the most important parts of data processing[28]. Although algorithms of data reduction have been developed, they are only for several specified instruments. The algorithm of data reduction should be improved for the instruments in China Spallation Neutron Source.

In this study, reduction algorithms for TOF-SANS with China Spallation Neutron Source NeXus event data are implemented in the Mantid project[29]. Based on the Mantid framework, every step of data reduction on the scattering function was extracted and analyzed, and the optimized reduction process was determined.

## Virtual neutron experiment

### Instrument setting

Figure 1 shows the TOF-SANS instrument used in the virtual experiment. Typically, it consists of one piece of guide, two apertures, three monitors, one beamstop, and one detector array. This spallation neutron source is selected with a repetition frequency of 25 Hz[30]. In order to increase the neutron flux and reduce simulation time, the divergence of the source was calculated using the distance between the two slits and that between slit 2 and the detector[31]. The distance between the sample and the detector is 6 m, equal to that between the two slits. The instrument parameters used in the virtual neutron experiment are listed in Table 1, and the spectrum of neutrons from the guide used in the simulation is shown in Fig. 2.

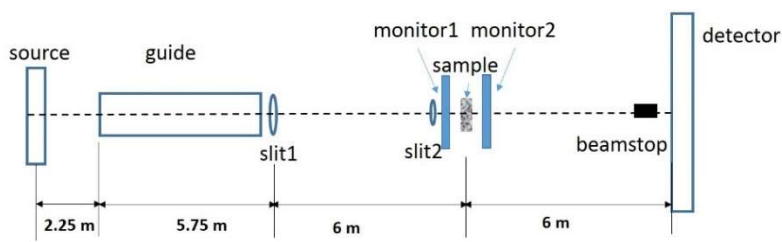

**Figure 1**. (Color online) Schematic diagram of the typical TOF-SANS instrument used in the virtual neutron experiment (not drawn to scale).

**Table 1**. Instrument parameters used in the virtual neutron experiment.

| Parameter | Value |
| --- | --- |
| Source frequency | 25 Hz |
| Source size | 0.04×0.04 m² |
| Distance from source to guide | 2.25 m |
| Guide length | 5.75 m |
| Radius of slit 1 | 0.016 m |
| Radius of slit 2 | 0.008 m |
| Distance between slit 1 and slit 2 | 6 m |
| Distance from sample to detector | 6 m |

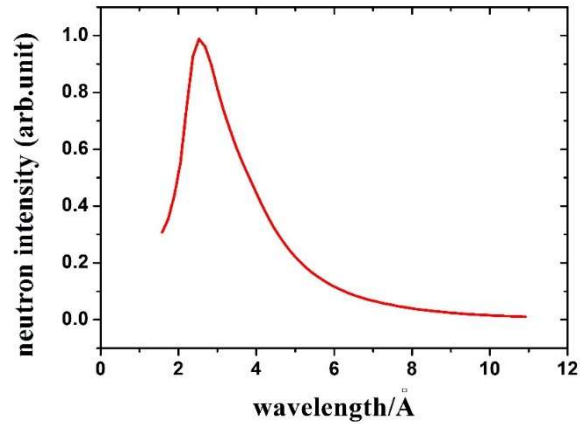

**Figure 2**. (Color online) Spectrum of neutrons from the guide used in the simulation.

**Detector simulation**

Detector simulation and digitization are implemented to generate detector output signals. In this study, 800 × 800 mm² ³He position-sensitive detectors were simulated. For each ³He detector tube, the response of two ends was simulated to obtain the charge of signal pulses ($Q_A$/$Q_B$) and the TOF. When a neutron is captured by the tube at position $z$, the charges are generated and propagated along the anode wire to both ends. The ratio of the charges collected at the two ends of anode can be derived based on the charge division method[32]. The readout module is implemented to create event-based binary files.

A reconstruction algorithm is developed in the data processing framework to calculate the hit position $P$ according to the formula

$$P = L\frac{Q_A}{Q_A + Q_B}, \qquad (1)$$

where $L$ is the length of the $^3$He tube. $Q_A$ and $Q_B$ are the charges of the two ends. Then, a mapping algorithm is implemented to determine the pixel identification from the hit position, module identification, and bank identification. The pixel size in this paper is $8 \times 8$ mm$^2$.

In actual experiments, the background of the empty container is supposed to have a lower scattering intensity than that of the sample. Further, the scattering intensity $I(Q)$ curve of an ideal empty container is expected to be as smooth as possible. The $I(Q)$ curves of the sample (circle) and the empty container (triangle) in the simulation are shown in Fig. 3. The simulated results agree with the analytical ones in the study. The lowest scattering vector $Q$ is around 0.008 Å$^{-1}$, according to the SANS configuration in the paper. Simulated data below 0.008 Å$^{-1}$ are unreliable owing to the strong direct neutron beam.

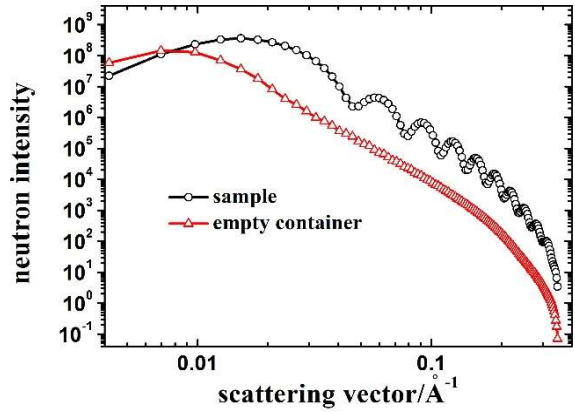

**Figure 3**. (Color online) $I(Q)$ curves of the sample (circle) and empty container (triangle) in the simulation.

In order to meet the requirements of data reduction, five values were measured by the virtual neutron experiment including sample scattering, sample transmission, empty container scattering, empty container transmission and direct beam.

## Data reduction

### Reduction principle

Data reduction, which involves a series of physical calibrations and corrections for experimental data, is performed to obtain instrument-independent data, which only reflects the intrinsic properties of the sample. In general, the measured data of a sample[33] has four scattering contributions including sample, empty container, solution and background. The experimental data of the sample, solvent, and empty container are expressed as

$$\begin{cases} I_{\exp}^{sample} = T_{ssc}(I_{cal}^{sample} + I_{cal}^{solvent} + I_{cal}^{can}) + I_{background} \\ I_{\exp}^{solvent} = T_{sc}(I_{cal}^{solvent} + I_{cal}^{can}) + I_{background} \\ I_{\exp}^{can} = T_c I_{cal}^{can} + I_{background} \end{cases}, \quad (2)$$

where the subscript *cal* indicates the theoretical calculation value. $T_{ssc}$, $T_{sc}$, and $T_c$ are the transmission coefficients of the sample, solvent, and empty container, respectively. $I_{background}$ represents the measured background data. Owing to the limitation in the accuracy of the Monte Carlo simulation, the scattering contributions of the background and solution are ignored in this study.

In addition, the scattering intensity obtained through experimental measurement is the sum of the intensities in all pixels of the neutron detector. Each record is significantly affected by detector performance, which is expressed as

$$I_{cal}(i, \lambda) = \frac{I_{\exp}(i, \lambda) M(i)}{\Omega(i)\varepsilon(i)I_{inc}(\lambda)\eta(\lambda)T(\lambda)}, \quad (3)$$

where $M(i)$ is the switch factor, masking the neutron events from the unexpected detector units; $\Omega(i)$

represents the contribution of the solid angle in the *i*-th pixel; $\varepsilon(i)$ is the detection efficiency in the *i*-th pixel; $\eta(\lambda)$ is the sensitivity of detector to neutrons with different wavelength; $I_{inc}(\lambda)$ indicates the incident neutron flux; and $T(\lambda)$ is the transmission coefficient.

Assuming $C_0 = \dfrac{M}{\Omega \varepsilon I_{inc} \eta}$,

$$I(\lambda) = \dfrac{I_{exp}^S C_0}{T^S} - \dfrac{I_{exp}^C C_0}{T^C}, \tag{4}$$

where superscript *S* represents the sample and *C* represents the container.

After determining the beamline center and masking detector and then converting units from TOF to wavelength, three major corrections are performed, including efficiency corrections for the detector and the monitor and transmission correction for the sample. Subsequently, the corrected data *I(λ)* are converted to *I(Q)*, and the scattering intensity contribution from the empty container is subtracted. In order to discuss the effect of these corrections in detail, the detector pixel No. 4495 (bright point in Fig. 4) is chosen for the result analysis.

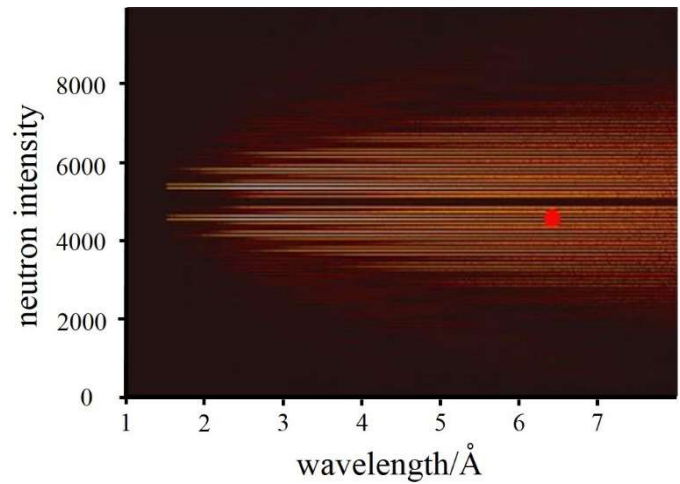

**Figure 4.** (Color online) Position of detector pixel No. 4495 (bright point).

**Determining the beam center**

Two-dimensional Gaussian distribution is applied to fit the incident neutron flux (see Fig. 5). When the beam center is determined, the position of each pixel of the area detector is also determined, including the relative position in both *x* and *y* directions, scattering angle $2\theta$ and azimuth $\Omega$. Therefore, mapping can be performed from TOF to the wavelength and/or the scattering vector $Q$.

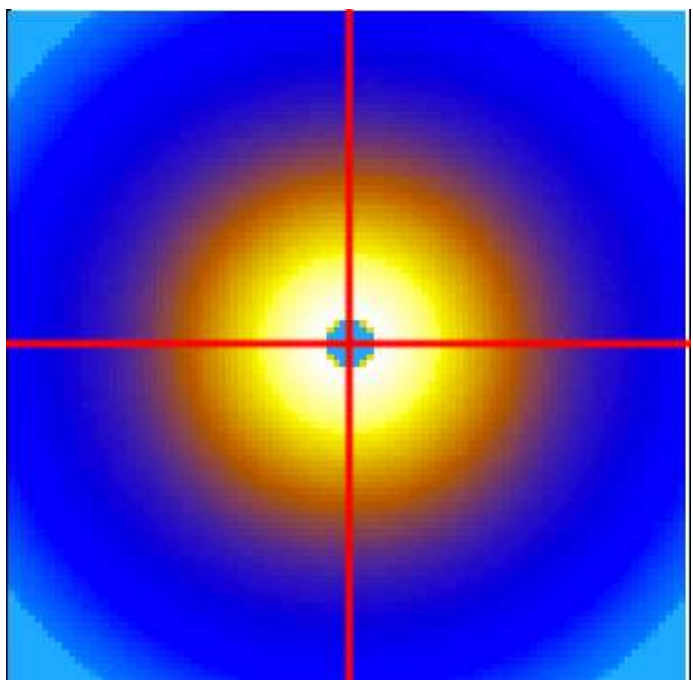

**Figure 5**. (Color online) Determination of beam center with two-dimensional Gaussian fitting.

**Detector masking**

In real experiments, there always exist bad pixels with high noise and zero or low counts in detectors. The data from those bad pixels should be ignored. In addition, a radial, angular, sectorial, or rectangular average is generally performed. The data must be truncated and sliced from the specific detected area. The pixels for the masking and averaging operations are defined in the mask file $M(i)$.

In the analysis, the size of the beamstop was set as $r = 0.032$ m, and the detector edge region was $r = 0.4$ m. As shown in the gray areas in Fig. 6, the detector pixels in the center and the edge are masked, and therefore, the radial average can be applied.

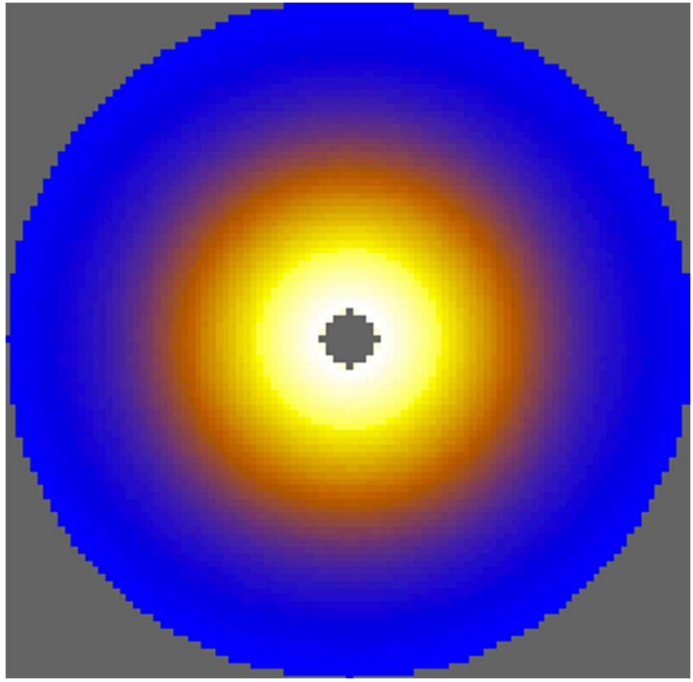

**Figure 6.** (Color online) Simulated scattering spot with masked pixels in the grey area.

**Mapping of TOF to wavelength**

The pulsed neutron is produced in the spallation neutron source. TOF technology is introduced to use a spectrum with a wide bandwidth. The neutron wavelength is determined by the TOF measured from the source to the detector. The conversion formula is

$$\lambda = \frac{ht}{mL}, \qquad (5)$$

where $t$ and $L$ are the time and distance of the flight from neutron source to the detector, respectively. $h$ is the Planck constant, and $m$ is the mass of the neutron.

Owing to the different positions of the incident monitor, transmission monitor, and detector units, the neutron wavelength is also different corresponding to the same flight time. However, all data processing must be performed on the same wavelength. As the resolution of the TOF diffractometer has an approximately logarithmic relationship with time ($\Delta\lambda/\lambda = $ constant), the TOF is rebinned in logarithmic format conventionally (see Fig. 7). After rebinning, the error in the neutron statistics is suppressed, and the consistency of data quality is maintained.

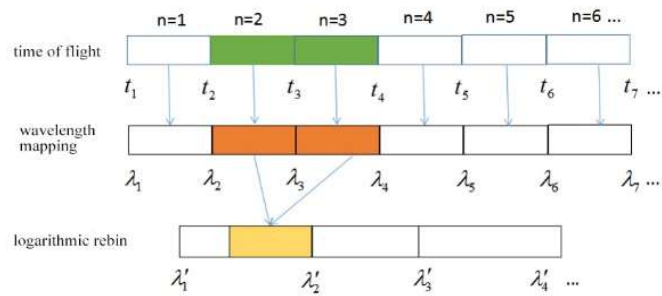

**Figure 7.** (Color online) Sketch of mapping from flight time to wavelength and data rebinning.

**Detector correction**

Scattering occurs on the Ewald sphere where area detectors are flat, and therefore, the counts of scattering neutrons recorded in the detector pixel are proportional to the solid angle. Solid angle correction is considered in data reduction. Each pixel covers different solid angles, which depends on the area of each pixel and the distance from the detector to the sample.

$$\Omega(2\theta) = \frac{p^2 \cos^3(2\theta)}{D(2\theta=0)}, \tag{6}$$

where $p$ represents the pixel size, $D$ is the distance from the sample center to the detector center, and $2\theta$ is the scattering angle. Figure 8 shows the comparison of the neutron intensity before and after solid angle correction for detector pixel no. 4495. In TOF-SANS, the distance from the sample to the detector is relatively large to achieve a lower scattering vector $Q$. Assuming the constant $p^2/D = 1$, the contribution of the solid angle varies from 0.993 to 0.999, where it is ~0.998 at pixel no. 4495. The uniformity across all detectors is extremely high.

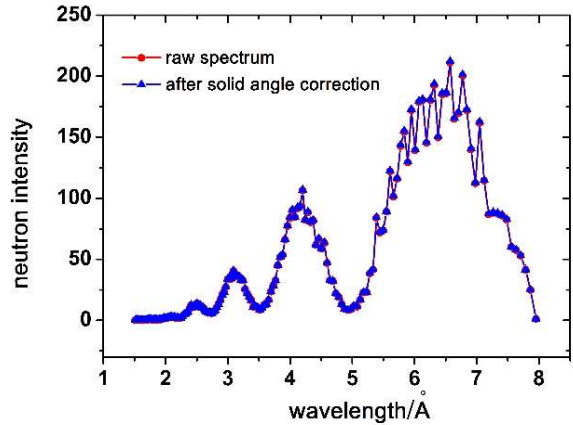

**Figure 8.** (Color online) Comparison of neutron intensity before and after solid angle correction of detector pixel no. 4495.

Detector efficiency[34] has several influence factors, including the transmittance of the incident particles, type of the working gas and absorption rate of the rays, and type and energy of the incident particles. Because of the spatial inhomogeneity in the area detector, efficiency correction is required to ensure the accuracy of the data.

The so-called flood data are generated from an isotropic scattering in the experiment, which takes into account the solid angle and detector efficiency. In this paper, the simulated flood data are shown in Fig. 9, which are normalized first by

$$S(x,y) = \frac{I^j_{flood}(x,y)}{1/N_{pixels} \Sigma_j I^j_{flood}(x,y)}, \qquad (7)$$

where $I^j_{flood}(x,y)$ is the neutron counts collected from the *j*-th pixel and $N_{pixel}$ is the number of detector pixels. The efficiency correction formula is

$$I_{cal} = I_{exp} / S(x, y). \tag{8}$$

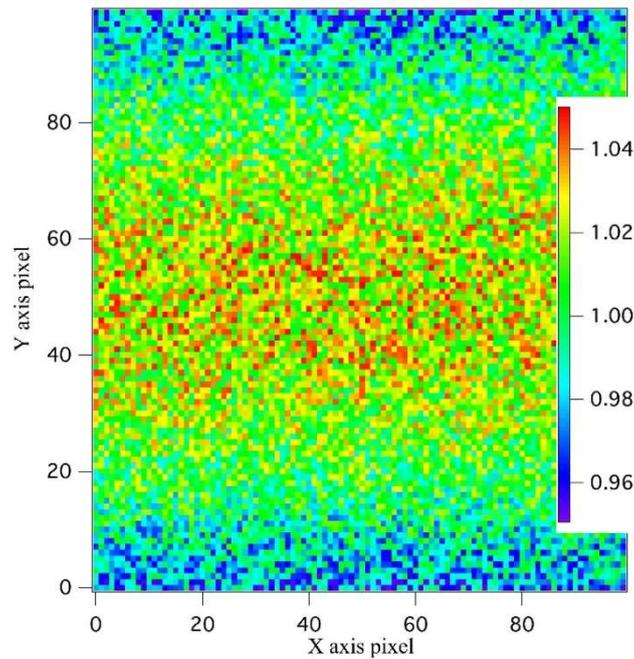

**Figure 9.** (Color online) Simulated flood data across all the detector pixels.

In the simulation, an ideal $^3$He tube is adopted, where the efficiency from the center to edge of the tube does not change significantly. The reduced efficiency is close to 1; therefore, the effect of efficiency on the result was not obvious (see Fig. 10). In practice, the efficiency distribution of the detector could be bad, and this could cause a remarkable deviation.

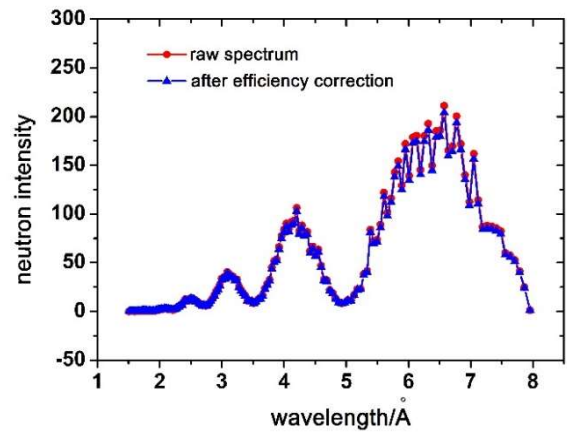

**Figure 10.** (Color online) Comparison of neutron intensity before and after efficiency correction of detector pixel no. 4495.

A detector's capability to collect a neutron is different for neutrons with different wavelength. In general, a neutron with long wavelength is more likely to be recorded by a detector. In the simulation, the neutron sensitivity of the monitor has an exponential dependence on neutron wavelength (see Fig. 11). Figure 12 shows a comparison of the results before and after the correction. The experimentally measured spectrum is found to have a very large deviation with the real incident neutron flux distribution.

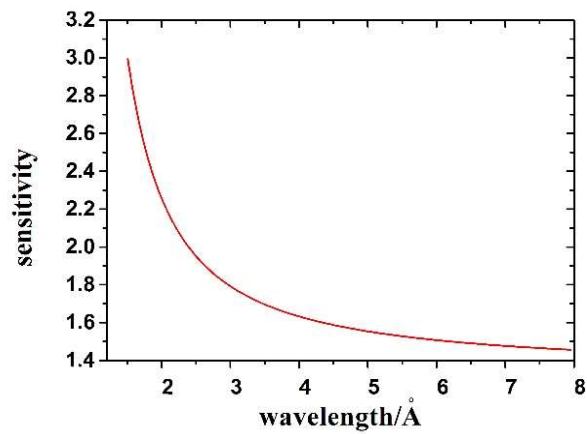

**Figure 11**. (Color online) Neutron sensitivity of the monitor versus wavelength.

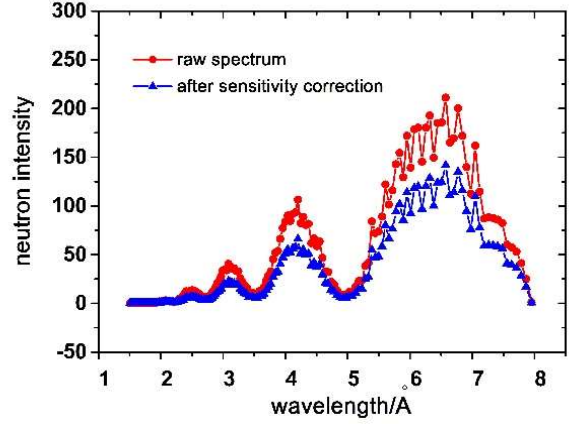

**Figure 12.** (Color online) Comparison of neutron intensity before and after sensitivity correction of detector pixel no. 4495.

**Transmission correction**

Most interactions between the neutron and the materials involve scattering and absorption. The other neutrons will transmit the sample to the beamstop or the detector. In order to decrease the effect of multiple scattering in the SANS experiment, the transmittance of the sample is set to 90% in the simulation, which is defined as the ratio of the transmission and the incident beam.

$$T(\lambda) = I_{trans}(\lambda) / I_{inc}(\lambda) \tag{9}$$

Considering the difference in the efficiency of the incident and transmission monitor in the real experiment, the direct beam experiment was simulated, in which the sample is empty, and the neutron counts recorded in the two monitors were used to correct the monitor efficiency. According to the analysis above, the formula for calculating the transmittance is

$$T = \frac{\varepsilon_{trans} I_{trans}^{sample}}{\varepsilon_{inc} I_{inc}^{sample}} \Big/ \frac{\varepsilon_{trans} I_{trans}^{direct}}{\varepsilon_{inc} I_{inc}^{direct}} = \frac{I_{trans}^{sample}}{I_{inc}^{sample}} \Big/ \frac{I_{trans}^{direct}}{I_{inc}^{direct}}, \tag{10}$$

where subscripts *inc* and *trans* are short for incident and transmission, respectively. $\varepsilon$ is the monitor efficiency. As can be seen, the differences between the incident and transmission monitor efficiency are counteracted after direct beam correction.

The transmission correction is also dependent on the scattering angle of the detector pixel, and the correction formula is

$$I_{cal}(\lambda) = \frac{I_{\exp}(\lambda)}{T^{[1+1/\cos(2\theta)]/2}} . \tag{11}$$

Figure 13 shows the comparison of neutron intensity before and after transmission correction. It is clear that the transmission has a large effect on scattering intensity. In practice, the transmission coefficient could appear to have a complicated dependence on the neutron wavelength. For the data reduction of SANS, transmission correction must be considered carefully.

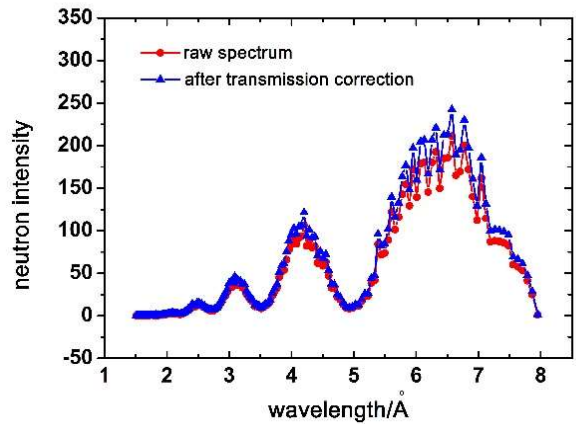

**Figure 13.** (Color online) Comparison of neutron intensity before and after transmission correction.

**Normalization of incident neutron**

One of important steps of TOF technology is to characterize the flux of the incident neutron with continuous wavelength. Owing to the spectrum distribution with the wavelength, normalization is implemented with the incident flux from the monitor just before the sample. Figure 14 illustrates the comparison of neutron intensity before and after normalization.

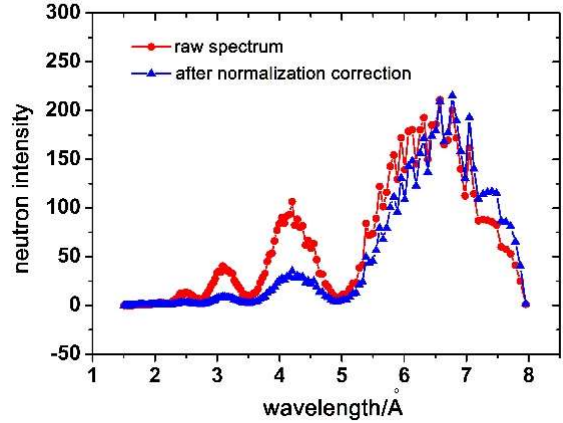

**Figure 14.** (Color online) Comparison of neutron intensity before and after normalization.

**Conversion from $I(\lambda)$ to $I(Q)$ considering gravity correction**

The plan is to reduce data at a pixel level first and then average over different wavelengths. The critical procedure in TOF-SANS data reduction is the conversion from wavelength to scattering vector. The scattering vector $Q$, as a function of wavelength, is

$$Q = \frac{4\pi}{\lambda}\sin(\theta), \qquad (12)$$

where $2\theta$ is the scattering angle. During neutron transportation from the source to the detector, a slow neutron is affected by gravity, and the calculated scattering vector deviates from the correct one. The actual position of the neutron with wavelength $\lambda$ recorded in the detector pixel is calculated as

$$\theta = \frac{1}{2}\arcsin\left(\frac{\sqrt{x^2 + (y + \frac{gm^2}{2h^2}\lambda^2 L_2^2)^2}}{L_2^2}\right), \qquad (13)$$

where $L_2$ is the distance from the sample to the detector. The gravity correction is necessary when a very slow neutron is adopted to approach a smaller scattering vector $Q$.

**Background subtraction**

Typical SANS data is contaminated with the scattering from containers, solution, and other sources. All background around the sample must be subtracted. As the simulation of the solution and the background are ignored in the virtual experiment, the contribution only from the container is subtracted using

$$I_{cal}^{sample} = T_{sample} I_{\exp}^{sample} - T_{can} I_{\exp}^{can}. \tag{14}$$

**Results and discussions**

Figure 15 shows a comparison of three scattering intensities, including theoretical calculation, virtual neutron experimental simulation, and the reduction in this study.

The resolution of the instrument was not considered in the theoretical calculation. The smearing effect [35] exists in both the simulation of the virtual neutron experiment and data reduction, and the trough of the wave weakened. In addition, the resolution of $\Delta Q$ in scattering intensity decreased with an increase in scattering vector $Q$. Further, the peak decreased in the large scattering vector $Q$ area.

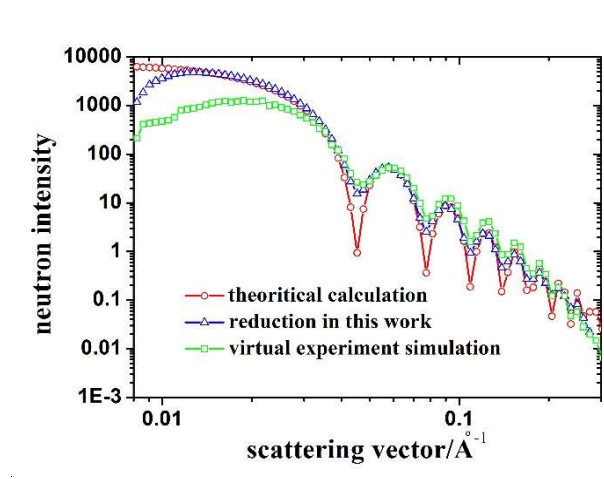

**Figure 15**. (Color online) Scattering intensities from theoretical calculation (circle and line), virtual neutron experimental simulation (square and line), and the reduction in this study (triangle and line).

It can be noticed that intensity reduced by our study has a large deviation with the theoretical value below a scattering vector $Q$ of 0.01 Å$^{-1}$. As mentioned in Section virtual neutron experiment, the minimum scattering vector $Q$ of the experiment is 0.008 Å$^{-1}$, with the bandwidth of the incident neutron of 1.5–7 Å. In fact, it is very difficult to obtain the accurate scattering intensity near the minimum scattering vector $Q$ in the real experiment. The data in the small scattering vector $Q$ is collected from the detector very close to the incident beam. It is inclined to be influenced by the strong direct beam, bad divergence, low spatial resolution, and high background. At the same time, the flux of the slow neutron is significantly low in the spallation neutron source, which leads to large statistical errors.

The reduction data matches well with the theoretical curve when the scattering vector $Q$ is in the range of 0.01–0.2 Å$^{-1}$. The crest and trough positions are consistent with the virtual experiments owing to the 100-nm-size spherical particles. However, as a large number of statistical errors exist in the Monte Carlo method used in the virtual neutron experiment, neither the calculation of transmission nor the simulation of background is sufficiently accurate. Although some improvements are required for the virtual neutron experiment performed in this study, the data reduction algorithm mentioned above covers all corrections of raw data for TOF-SANS.

## Conclusion

A virtual neutron experiment was used to simulate a general TOF-SANS instrument in this study, and the reduction was performed based on Monte Carlo scattering data. The effect of each reduction algorithm was analyzed. After a series of complete corrections, the processed scattering intensity from the TOF-SANS experiment is consistent with the theoretical calculation value and the virtual neutron experiment value, which demonstrates the accuracy of the simulation and the reduction. The deviation in the result mainly originated from the precision limit of the Monte Carlo method and the incomplete simulation of the instrument.

Although reduction algorithms for TOF-SANS depend on the specific neutron source and instrument configuration, the detailed study of data reduction can help design and optimize the neutron instrument. More importantly, this study helps to provide and understand high-quality data, and to bridge the gap between the SANS instrument and materials research.

In 2018, China Spallation Neutron Source will be completed, and the first TOF-SANS of China will be operational. Our work will lay a foundation for the analysis of data measured by SANS at China Spallation Neutron Source.

# Reference

[1]  Windsor, C. An introduction to small-angle neutron scattering. J. Appl. Crystallogr. 1988, 21, 582-588.

[2]  Melnichenko, Y. B.; Wignall, G. D. Small-angle neutron scattering in materials science: recent practical applications. J. Appl. Phys. 2007, 102, 021101.

[3]  Penfold, J.; Thomas, R K. Neutron reflectivity and small angle neutron scattering: an introduction and perspective on recent progress. Curr. Opin. Colloid Interface Sci. 2014, 19, 198-206.

[4]  MikhailL, V. A.; Viktor, A. Small-angle neutron scattering in structure research of magnetic fluids. Phys-Usp. 2010, 53, 971.

[5]  Gomes, S. R.; Margaça, F. M. A.; Miranda-Salvado, I. M.; Falcao, A. N.; Almasy, L.; Teixeira, J. SANS investigation of PDMS hybrid materials prepared by gamma-irradiation. Nucl. Instrum. Methods Phys. Res., Sect. B. 2008, 266, 5166-5170.

[6]  Pan, J.; Heberle, F. A.; Petruzielo, R. S.; Katsaras, J. Using small-angle neutron scattering to detect nanoscopic lipid domains. Chem. Phys. Lipids. 2013, 170–171, 19-32.

[7]  Andreas, M. Magnetic small-angle neutron scattering of bulk ferromagnets. J. Phys.: Condens. Matter. 2014, 26, 383201.

[8]  Rharbi, Y.; Yousfi, M.; Porcar, L.; Nawaz, Q. Methods for probing the long-range dynamic of confined polymers in nanoparticles using small-angle neutron scattering. Can. J. Chem. 2010, 88, 288-297.

[9]  Strobl, M.; Steitz, R.; Kreuzer, M.; Rose, M.; Herrlich, H.; Mezei, F.; Grunze, M.; Dahint, R. BioRef: a versatile time-of-flight reflectometer for soft matter applications at Helmholtz–Zentrum

Berlin. Rev. Sci. Instrum. 2011, 82, 055101.

[10] Heenan, R. K.; Penfold, J.; King, S. M. SANS at pulsed neutron sources: present and future prospectst. J. Appl. Crystallogr. 1997, 30, 1140-1147.

[11] Seeger, P. A.; Hjelm, R. P. Small-angle neutron scattering at pulsed spallation sources. J. Appl. Crystallogr. 1991, 24, 467-478.

[12] Kynde, S.; Hewitt-Klenø, K.; Nagy, G.; Mortensen, K.; Lefmann, K. A compact time-of-flight SANS instrument optimised for measurements of small sample volumes at the European Spallation Source. Nucl. Instrum. Methods. Phys. Res., Sect. A. 2014, 764, 133-141.

[13] Kaune, G.; Haese-Seiller, M.; Kampmann, R.; Moulin, J. F.; Zhong, Q. TOF-GISANS Investigation of Polymer Infiltration in Mesoporous $TiO_2$ Films for Photovoltaic Applications. J. Polym. Sci., Part B: Polym. Phys., 2010, 48, 1628-1635.

[14] Muller-Buschbaum, P.; Metwalli, E.; Moulin, J. F.; Kudryashov, V.; Haese-Seiller, M.; Kampmann, R. Time of flight grazing incidence small angle neutron scattering. Eur. Phys. J. Special Topics 2009, 167, 107-112.

[15] Bentley, P. M.; Andersen, K. H. Optimization of focusing neutronic devices using artificial intelligence techniques. J. Appl. Crystallogr. 2009, 42, 217-224.

[16] Lefmann, K.; Willendrup, P. K.; Udby, L.; Lebech, B.; Mortensen, K.; Birk, J.O.; Klenø, K.; Knudsen, E.; Christiansen, P.; Saroun, J.; Kulda, J.; Filges, U.; Konnecke, M.; Tregenna-Piggott, P.; Peters, J.; Lieutenant, K.; Zsigmond, G.; Bentley, P.; Farhi, E. Virtual experiments: the ultimate aim of neutron ray-tracing simulations. J. Neutron. Res. 2008, 16, 97-111.

[17] Udby, L.; Willendrup, P. K.; Knudsen, E.; Niedermayer, Ch.; Filges, U.; Christensen, N. B.; Farhi, E.; Wells, B. O.; Lefmann, K. Analysing neutron scattering data using McStas virtual experiments.

Nucl. Instrum. Methods. Phys. Res., Sect. A. 2011, 634, S138-S143.

[18] Paul, A. Wavelength resolution options for a time-of-flight reflectometer using VITESS code of simulation. Nucl. Instrum. Methods. Phys. Res., Sect. A 2011, 646, 158-166.

[19] Zendler, C.; Lieutenant, K.; Nekrassov, D.; M, Fromme. VITESS 3-virtual instrumentation tool for the European Spallation Source. J. Phys. 2014, 528, 012036.

[20] Uca, O.; Ohms, C. Monte Carlo simulations of the SANS instrument in Petten. Physica. B. 2008, 403, 3892-3895.

[21] Jaksch, S.; Martin-Rodriguez, D.; Ostermann, A.; Jestin, J.; Duarte Pinto, S.; Bouwman, W. G.; Uher, J.; Engels, R. Frielinghaus, H. Concept for a time-of-flight small angle neutron scattering instrument at the European Spallation Source. Nucl. Instrum. Methods. Phys. Res., Sect. A. 2014, 762, 22-30.

[22] Tian, H. L.; Zhang, J. R.; Yan, L. L.; Tang, M.; Hu, L.; Zhao, D. X.; Qiu, Y. X.; Zhang, H. Y.; Zhuang, J.; Du, R. Distributed data processing and analysis environment for neutron scattering experiments at CSNS. Nucl. Instrum. Methods. Phys. Res., Sect. A. 2016, 834, 24-29.

[23] Konnecke, M.; Akeroyd, F. A.; Bernstein, H. J.; Brewster, A. S.; Campbell, S. I.; Clausen, B.; Cottrell, S.; Hoffmann, J. U.; Jemian, P. R.; Mannicke, D.; Osborn, R.; Peterson, P. F.; Richter, T.; Suzuki, J.; Watts, B.; Wintersberger, E.; Wuttke, J. The NeXus data format. J. Appl. Crystallogr. 2015, 48, 301-305.

[24] Zhao, J. K. Data processing for the SNS EQ-SANS diffractometer. Nucl. Instrum. Methods. Phys. Res., Sect. A. 2011, 647, 107-111.

[25] Peterson, P. F.; Campbell, S. I.; Reuter, M. A.; Taylor, R. J.; Zikovsky, J. Event-based processing of neutron scattering data . Nucl. Instrum. Methods. Phys. Res., Sect. A. 2015, 803, 24-28.

[26] Kampman, R.; Haese-Seiller, M.; Kudryashov, V.; Nickel, B.; Daniel, C.; Fenzl, W.; Schreyer, A.; Sackmann, E.; Radler, J. Horizontal Tof-neutron reflectometer REFSANS ata FRM-II Munich/Germany: First tests and status. Phys. B., 2006, 385-386, 1161-1163.

[27] Marmotti, M.; Haese-Seiller, M.; Kampmann, R. Two-dimensional position-sensitive 3He-neutron detector for reflectometry and high-resolution diffractometry. Phys. B., 2000, 276-278, 210-211.

[28] Kline, S. R. Reduction and analysis of SANS and USANS data using IGOR Pro. J. Appl. Crystallogr. 2006, 39, 895-900.

[29] Arnold, O.; Bilheux, J. C.; Borreguero, J. M.; Buts, A.; Campbell, S. I.; Chapon, L.; Doucet, M.; Draper, N.; Ferraz-Leal, R.; Gigg, M. A.; Lynch, V. E.; Markvardsen, A.; Mikkelson, D. J.; Mikkelson, R. L.; Miller, R.; Palmen, K.; Parker, P.; Passos, G.; Perring, T. G.; Peterson, P. F.; Ren, S.; Reuter, M. A.; Savici, A. T.; Taylor, J. W.; Taylor, R. J.; Tolchenov, R.; Zhou, W.; Zikovsky, J. Mantid—Data analysis and visualization package for neutron scattering and μ SR experiments. Nucl. Instrum. Methods. Phys. Res., Sect. A. 2014, 764, 156-166.

[30] Yin, W.; Liang, T. J.; Yu, Q. Z. Neutronics design for the coupled para-hydrogen moderator for CSNS. Nucl. Instrum. Methods. Phys. Res., Sect. A. 2011, 631, 105-110.

[31] Zuo, T. S.; Cheng, H.; Chen ,Y. B.; Wang, F. W. Development and prospects of Very Small Angle Neutron Scattering (VSANS) techniques. Chinese. Phys. C. 2016, 40, 076204.

[32] Radeka, V.; Rehak, P. Charge dividing mechanism on resistive electrode in position-sensitive detectors. IEEE Trans. Nucl. Sci. 1979, NS-26, 225-238.

[33] Wei, G. H.; Liu, X. F.; Li, T. F.; Zhang, L.; Wang, Y.; Wang, H. L. Small angle neutron scattering experiment and raw data reduction (in Chinese). Nucl. Tech. 2010, 33, 253-257.

[34]Chen, L.; Peng, M.; Sun, L. W.; Chen, B. The calibration and test methods of position sensitive detector for small angle neutron sacttering spectrometer (in Chinese). Nucl. Electron. Detection. Tech. 2012, 32, 656-659.

[35]Barker, J. G.; Pedersen, J. S. Instrumental smearing effects in radially symmetric small-angle neutron scattering by numerical and analytical methods. J. Appl. Crystallogr. 1995, 28, 105-114.